\documentstyle[epsfig,aps]{revtex}

\def\be{\begin{eqnarray}}
\def\ee{\end{eqnarray}}
\def\nn{\nonumber\\ }
\def\Tr{{\rm Tr}\;}

\def\uv{\underline{v}}
\def\ua{\underline{a}}

\def\T{{\rm T}}

\def\hS{\hat {\cal S}}

\def\J{{\rm\bf J}}
\def\K{{\rm\bf K}}

\def\V{{\rm\bf V}}

\def\fm{{\rm fm}}

\def\j{{\rm\bf j}}

\def\one{{\rm\bf 1}}

\def\bi{\begin{itemize}}
\def\ei{\end{itemize}}
\def\tI{\tilde{\cal I}}

\begin{document}

\title{Master Formulae Approach to Photon Fusion Reactions}

\author{C.-H. Lee$^a$, H. Yamagishi$^b$ and I. Zahed$^a$}
\address{a) Department of Physics \& Astronomy, SUNY at Stony Brook,
Stony Brook, NY 11794, USA;\\
b) 4 Chome 11-16-502, Shimomeguro, Meguro, Tokyo, 153 Japan.}

\maketitle

\begin{abstract}
We analyze the $\gamma\gamma\rightarrow \pi\pi, KK, \eta\eta, \pi\eta$
reactions through $\sqrt{s}=2$ GeV, using the master formula approach to
QCD with three flavors.
In this approach, the constraints of broken chiral symmetry, unitarity
and crossing symmetry are manifest in all channels. The pertinent vacuum 
correlators are analyzed at tree level using straightforward resonance 
saturation methods. A one-loop chiral power counting analysis at treshold
is also carried out and compared to standard chiral perturbation theory. 
Our results are in overall agreement with the existing data in all channels.
We predict the strange meson polarizabilities and a very small cross section
for $\gamma\gamma\rightarrow\eta\eta$.

\end{abstract}        

\section{Introduction}
\label{sec:1}

There is a wealth of empirical information regarding photon fusion
reactions to two mesons both at threshold and 
above~\cite{TPC,MARK,DESY,KEK,DORIS,SAN,KKpm1,TASSO,CELLO,pieta}.
At low energy these reactions provide stringent constrainsts on our
understanding of broken chiral symmetry and the way mesons and 
photons interact. At higher energy they reveal a variety of resonance
structure that reflects on the importance of final state correlations
and unitarity in strong interaction physics.

Some of these reactions have been analyzed using chiral perturbation
theory~\cite{dono93}, dispersion relations~\cite{DISPERSION} and also 
effective models~\cite{EFFECTIVE}. One-loop chiral perturbation theory
~\cite{ONE} does well in the charged channels, but yields results that 
are at odd with the data in the chargeless sector, suggesting that
important correlations are at work in the final states. Some of these
shortcomings have been removed by a recent two-loop calculations~\cite{TWO}
and the help of few parameters that are fixed by resonance saturation. 
The results are overall in agreement with an early dispersion analysis
for $\gamma\gamma\rightarrow \pi^0\pi^0$~\cite{DISPERSION}. Effective
models using aspects of chiral symmetry and s-channel unitarisation
have revealed the importance of final state interactions in most of 
these reactions~\cite{EFFECTIVE,OSET}.

Recently, a global and unified understanding of broken chiral symmetry 
was reached in the form of a master formula for the extended S-matrix
\cite{YAZA}. A number of reaction processes involving two light quarks
were worked out and shown to be interdependent beyond threshold. The 
approach embodies the essentials of broken chiral symmetry, unitarity
and crossing symmetry to all orders in the external momenta. By power
counting it agrees with standard chiral perturbation theory in the
threshold region. It is
flexible enough to be used in conjunction with dispersion analysis or
resonance saturation techniques to allow for a simple understanding 
of resonance effects and final state interactions beyond threshold.

In this paper, we would like to give a global understanding of most of
the fusion reaction processes using the master formula approach to broken 
chiral symmetry including the effects of strangeness. The present work 
confirms and extends the original analysis in the two flavour case
\cite{CZ95}. In section~\ref{sec:2}, we 
introduce our conventions for the fusion reaction processes, and discuss the 
essentials of the T-matrix amplitudes. In section~\ref{sec:3}, 
we give the main result
for the fusion reaction processes as expected from the master formula approach
to QCD with three flavors. The importance of s-channel scalar correlations 
is immediately unravelled. In section~\ref{sec:4}, 
we analyze the general result 
in chiral power counting and compare to one-loop chiral perturbation theory
with strangeness. In section~\ref{sec:5}, 
we analyze the master formula result beyond 
threshold by using resonance saturation methods. 
In section~\ref{sec:6}, we discuss
briefly the meson polarizabilities in our case. In section~\ref{sec:7}, 
a detailed 
numerical analysis of our results is made and compared to presently available
data. We predict a small cross section for $\gamma\gamma\rightarrow 
\eta\eta$. Our conclusions are in section~\ref{sec:8}. 
Some calculational details are
given in three Appendices.

\section{Generalities}
\label{sec:2}

\subsection{Conventions}

We will consider generically the reactions 
$\gamma^c(q_1) \gamma^d(q_2)\rightarrow \pi^a(k_1)\pi^b(k_2)$
with $a,b=1\sim 8$ and $c,d=3,8$ for the light mesonic octet.
The photon polarizations are chosen in the gauge 
$\epsilon_\mu(q_i) q_j^\mu =0$ with $i,j=1,2$.
Throughout, the Mandelstam variables are given by
     \be 
     s &=& (q_1+q_2)^2 = 2q_1\cdot q_2 \nonumber\\
     t &=& (q_1-k_1)^2 = k_1^2 - 2 q_1\cdot k_1 \nonumber\\
     u &=& (q_1-k_2)^2 = k_2^2 - 2 q_1\cdot k_2 \ .
     \ee
and both the photons and the mesons are on-shell, $q_i^2=0$ and
$k_i^2 = m_{a}^2$. Our convention for the electromagnetic current 
is standard
     \be
     \J_\mu^{em} = \bar q\gamma_\mu \left(
      \frac 1 2 \lambda_3 +\frac{1}{2 \sqrt 3}\lambda_8 \right) q 
      = \V_\mu^3 +\frac{1}{\sqrt 3} \V_\mu^8 \ ,
     \ee      
so that the photon isospin indices are only 3 and 8. This will be assumed
throughout.

\subsection{Helicity Amplitudes}

The T-matrix for the fusion process 
$\gamma(q_1)\gamma(q_2)\rightarrow \pi^a(k_1)\pi^b(k_2)$,  
will be defined as~\cite{TWO} 
     \be
     {}_{\rm out}\langle \pi^a(k_1)\pi^b(k_2) 
         |\gamma(q_1)\gamma(q_2)\rangle_{\rm in}
     =i (2\pi)^4\delta^4 (P_f-P_i) {\cal T}^{ab}
     \ee
with 
    \be
    {\cal T} = e^2 \epsilon_1^\mu \epsilon_2^\nu V_{\mu\nu}^{ab} \ .
    \ee
The photons are transverse, that is $\epsilon_i\cdot q_j=0$, hence
    \be
    V_{\mu\nu}  = A(s,t,u) T_{1\mu\nu} + B(s,t,u) T_{2\mu\nu} 
    \label{tmat}
    \ee
with the invariant tensors
    \be
    T_{1\mu\nu} = \frac 12 s g_{\mu\nu}- q_{1\mu}q_{2\nu} \qquad\qquad
    T_{2\mu\nu} =  2 s (k_1-k_2)_\mu (k_1-k_2)_\nu - \nu^2 g_{\mu\nu}
    \ee
and $\nu=(t-u)$. As a result, the T-matrix reads
    \be
    {\cal T} &=& e^2 \left({A(s,t,u)} \,s/2 - \nu^2 B(s,t,u)\right)
                   \epsilon_1\cdot\epsilon_2 
         - e^2 8 s B(s,t,u) \epsilon_1\cdot k_1 \epsilon_2\cdot k_2
    \nn
       &=& -2 e^2 \epsilon_1\cdot\epsilon_2 (\one - {\cal X})
           -e^2 (\epsilon_1\cdot k_1) (\epsilon_2\cdot k_2)
             8 s ( B_0 +{\cal Y})
    \ee
with $\one$ and $B_0$  defined as 
    \be
     \one &=& \left\{ \begin{array}{cl} 1 & {\rm for}\; \pi^\pm, K^\pm \\
                       0 & {\rm for}\; \pi^0,K^0,\bar K^0,\eta
                   \end{array} \right. \nn
    B_0 &=& \one
    \frac{1}{2s}\left(\frac{1}{t-m_\pi^2}+\frac{1}{u-m_\pi^2}\right) \ .
    \ee
The corresponding helicity amplitudes are~\cite{TWO}
    \be
    H_{++}^{ab} &=& A^{ab} + 2( (m_a+ m_b)^2 -s) B^{ab} \nn
    H_{+-}^{ab} &=& \frac{8 (m_a^2 m_b^2-tu)}{s} B^{ab} \ .
    \ee

\subsection{Polarizabilities}

The differential cross section for unpolarized photons to two mesons in the
center-of-mass system is
    \be
    \frac{d\sigma^{\gamma\gamma\rightarrow \pi^a\pi^b}}{d\Omega}
      &=& f_{ab} \frac{\alpha^2 s}{32}\beta^{ab}(s)
           \left(|H_{++}|^2+|H_{+-}|^2\right)
    \nn
      &=& f_{ab} \frac{\alpha^2 }{4s}\beta^{ab}(s)
     \left( \left| {\cal B}+\frac{m_\pi}{2\alpha} s\alpha^{ab}_\pi(s)\right|^2
     +\left| {\cal B}^\prime+\frac{m_\pi}{2\alpha} s\alpha^{ab}_\pi(s)\right|^2
       \right) \ ,
    \ee
with the degeneracy factor 
    \be
    f_{ab} = \left\{
    \begin{array}{cl}
         1/2 & {\rm for}\;\; \pi^0\pi^0, \eta\eta \\
         1   & {\rm for}\;\; {\rm other \;\; processes} 
    \end{array}\right. \ .
    \ee
The expressions for ${\cal B}$, ${\cal B}^\prime$ and the
polarizabilities $\alpha_\pi^{ab}$ are 
    \be
    {\cal B} &=& \one
    \left(-1 +\frac{2s m_\pi^2}{(t-m_\pi^2)(u-m_\pi^2)} \right)
    \nn
    {\cal B}^\prime &=& \one +4(m_\pi^4-tu) {\cal Y} \nn
    \frac{m_\pi}{2\alpha} s\alpha_\pi^\pm(s) &=&
      -{\cal X}-\frac{s(4m_\pi^2-s)+4 (m_\pi^4-tu)}{2}{\cal Y}\ .
    \ee
The center of mass velocity for outgoing particles
$\beta^{ab}(s)$ will be defined as
    \be
    \beta^{ab}(s) = \sqrt{\left(1-\frac{(m_a+m_b)^2}{s}\right)
                          \left(1-\frac{(m_a-m_b)^2}{s}\right)} \ .
    \ee

\section{Master Formulae Result}
\label{sec:3}

The master formula approach to two flavours developed by two of 
us~\cite{YAZA} can be readily extended to three flavours~\cite{LYZ98}.
In short, the extended S-matrix with strangeness included obeys a new 
and linear master equation, that is emmenable to on-shell chiral
reduction formulas. The fusion reaction processes can be assessed as
discussed in ~\cite{YAZA,CZ95} for two flavours. 
The three flavour result is
     \be
     {\cal T}_1 &= & i \epsilon_1\cdot\epsilon_2
      \frac{E_a}{E_b} (f^{bci} f^{ida} +f^{bdi} f^{ica} )
     \nn
     && +
     i 4 \epsilon_1\cdot k_2 \epsilon_2\cdot k_1 \frac{E_a}{E_b}
     \left\{  \frac{f^{bci}f^{ida}}{u-m_i^2} 
     +  \frac{f^{bdi}f^{ica}}{t-m_i^2} \right\}
    \label{eq:T1}
     \\
     {\cal T}_2 &=& i \epsilon_1\cdot\epsilon_2 \frac{1}{E_a E_b} f^{bdi}f^{aci}
      \left\{\frac 23 K \frac{M_a}{m_a^2} - E_i^2 \right\}
     \nn
      && + \epsilon_1^\mu\epsilon_2^\nu
      k_2^\beta k_1^\alpha \frac{1}{E_a E_b}
     \int d^4z\int d^4y\int d^4x \,e^{ik_2\cdot x-i q_1\cdot y-i q_2\cdot z}
     \langle \,T^*\,\V_\nu^d(z)\V_\mu^c(y){\j_A}_\beta^b(x){\j_A}_\alpha^a(0)\rangle
     \nn
     && + i\epsilon_1^\mu\epsilon_2^\nu
     \frac 23\frac KC \delta^{ab} \frac{M_a}{E_a^2}
     \int d^4z\int d^4y\, e^{-iq_1\cdot y -i q_2\cdot z}
      \langle\,T^*\, \V_\nu^d(z)\V_\mu^c(y)\sigma^0(0)\rangle
     \nn
     && -i \epsilon_1^\mu\epsilon_2^\nu d^{abh} 
      \frac{M_b}{E_a E_b} \frac{E_h m_h^2}{M_h}
     \int d^4z\int d^4y\, e^{-iq_1\cdot y -i q_2\cdot z}
      \langle\,T^*\, \V_\nu^d(z)\V_\mu^c(y)\sigma^h(0)\rangle \ ,
    \label{eq:T2}
     \ee
where ${\cal T}_1$ summarizes the Born contributions to the charged
mesons, and ${\cal T}_2$ the rest after two chiral reductions of the
external meson states. (\ref{eq:T2}) constitutes our basic identity.
It shows that the fusion reaction is related to the vacuum correlators
${\bf V}{\bf V}{\bf j}{\bf j}$ and ${\bf V}{\bf V}\sigma$ modulo Born 
terms. Quantum numbers and G-parity imply that the scalars dominate the 
final state interactions in the s-channel. This point will become clearer in 
the resonance saturation analysis. What is remarkable in (\ref{eq:T2})
is that the final state scalar correlations are driven by the symmetry
breaking effects in QCD.

In (\ref{eq:T2}) 
the isovector current ${\bf V}$ and the one-pion reduced iso-axial current 
${\bf j}_A$ are given by
\be
 \V^a_\mu = \bar q \gamma_\mu \frac{\lambda^a}{2} q\ , \;\;\;\;
 {\j_A^a}_\mu = \bar q \gamma_\mu \frac{\lambda^a}{2} \gamma_5 q\  
  + \left(\frac{M}{m_p^2}\right)^{ab} 
   \partial_\mu (\bar q i\gamma_5\lambda^b q ) 
\ .
\ee
The mesons weak-decay 
constants and masses are
     \be
     E_{1\sim 8} & \equiv & (f_\pi,f_\pi,f_\pi,f_K,f_K,f_K,f_K,f_\eta)
     \nonumber\\
     m_{1\sim 8} &\equiv &  (m_\pi,m_\pi,m_\pi,m_K,m_K,m_K,m_K,m_\eta)
     \ee
with  $f_\pi=93$ MeV, $f_K=115$ MeV and $f_\eta=123$ MeV.
The current mass matrix is chosen as
    \be
     M_{1\sim 8} \equiv \left(\hat m,\hat m,\hat m, \frac{\hat m+m_s}{2},
      \frac{\hat m+m_s}{2}, \frac{\hat m+m_s}{2},
      \frac{\hat m+m_s}{2}, \frac{\hat m+2 m_s}{3} \right) 
     \ee
with $\hat m=9$ MeV and $m_s=175$ MeV for some running scale.
Since the $M$'s appear in RGE invariant combinations, the effects of 
the running scale is small in the range of energies probed by the 
fusion reaction processes we will be considering.
The scalar densities are
     \be
     \sigma^0 &=& \frac CK \bar q q +C \nn
     \sigma^h &=& -\frac{M_a}{E_a m_a^2} \bar q\lambda^a q \ .
     \label{eq:sigma}
     \ee
with two (arbitrary) constants. For two flavours, $C \rightarrow -f_\pi$ 
and $2 K/3  \rightarrow  f_\pi^2 m_\pi^2/\hat m$.

The contact term in ${\cal T}_2$ vanishes in the two-flavour case. It does 
not in the tree-flavour case and is to be reabsorbed in the pertinent
counterterm generated by the three and four point functions in (\ref{eq:T2}). 
The Born terms involve only charged mesons. Their explicit form is
     \be
     {\cal T}_{\gamma\gamma\rightarrow\pi^+\pi^-}
      &=&
      -i 2 e^2 \epsilon_1\cdot\epsilon_2
      -i 4 e^2 \epsilon_1\cdot k_1 \epsilon_2\cdot k_2
      \left(\frac{1}{t-m_\pi^2}+\frac{1}{u-m_\pi^2}\right)
     \nn
     {\cal T}_{\gamma\gamma\rightarrow K^+ K^-}
      &=& -i 2 e^2 \epsilon_1\cdot\epsilon_2
      -i 4 e^2 \epsilon_1\cdot k_1 \epsilon_2\cdot k_2
      \left(\frac{1}{t-m_K^2}+\frac{1}{u-m_K^2}\right) \ .
     \ee
These tree results are consistent with all chiral models with minimal
coupling. To go beyond, we need to assess the effects of the three- and 
four-point functions in (\ref{eq:T2}). We will do this in two ways : at 
threshold by using power counting, and beyond threshold by using resonance
saturation methods.


\section{One Loop Result}
\label{sec:4}

The identity (\ref{eq:T2}) is a consequence of broken chiral symmetry 
in QCD, and any chiral approach that is consistent with QCD ought to
satisfy it. In this section we show how this identity can be analysed
near treshold using power counting in $1/E$. A simple comparison with 
the nonlinear sigma model shows that this is analogous to the loop 
expansion if $\phi={\bf V}, {\bf j}_A, \sigma$ are counted of order 
${\cal O}(1)$. Also $f_K^2-f_{\pi}^2$ and $f_{\eta}^2-f_{\pi}^2$ are
${\cal O}(1)$ rather than ${\cal O}(E)$ because of G-parity. 

Some details regarding the one-loop analysis are given in Appendix B.
The results for the various transition amplitudes are
     \be
     {\cal T}_{\gamma\gamma\rightarrow\pi^+\pi^-}
      &=&
       -i 2 e^2 k_1\cdot k_2 \frac{1}{f_\pi^2} 
       \left(\tI^\pi +\frac 12 \tI^K\right)
       -i2 e^2 \frac{m_\pi^2}{f_\pi^2} \tI^\pi 
       -i e^2\frac{m_K^2}{f_\pi^2}\frac{2\hat m}{\hat m+m_s} \tI^K
     \nn
     {\cal T}_{\gamma\gamma\rightarrow\pi^0\pi^0} &=&
      -i 2 e^2 k_1\cdot k_2 \frac{1}{f_\pi^2} 
      \left(2\tI^\pi +\frac 12 \tI^K\right)
      -i2 e^2 \frac{m_\pi^2}{f_\pi^2} \tI^\pi 
      -i e^2\frac{m_K^2}{f_\pi^2}\frac{2\hat m}{\hat m+m_s} \tI^K
     \nn
    {\cal T}_{\gamma\gamma\rightarrow K^+ K^-}
     &=& -i 2 e^2 k_1\cdot k_2 \frac{1}{f_K^2} 
     \left(\frac 12\tI^\pi +  \tI^K\right)
      -i e^2 \frac{m_\pi^2}{f_K^2} \frac{\hat m+m_s}{2\hat m}\tI^\pi 
      -i \frac 32 e^2\frac{m_K^2}{f_K^2} \tI^K
    \nn
    {\cal T}_{\gamma\gamma\rightarrow K^0 \bar K^0}
     &=& -i 2 e^2 k_1\cdot k_2 \frac{1}{f_K^2} 
     \left(\frac 12\tI^\pi + \frac 12 \tI^K\right)
      -i e^2 \frac{m_\pi^2}{f_K^2} \frac{\hat m+m_s}{2\hat m}\tI^\pi 
      -i \frac 32 e^2\frac{m_K^2}{f_K^2} \tI^K
    \nn
    {\cal T}_{\gamma\gamma\rightarrow \eta \eta}
     &=& -i 3 e^2 k_1\cdot k_2 \frac{1}{f_\eta^2} \tI^K
      -i \frac 23 e^2 \frac{m_\pi^2}{f_\eta^2} 
      \frac{\hat m+ 2 m_s}{3\hat m}\tI^\pi 
      -i \frac 53 e^2\frac{m_K^2}{f_\eta^2}
      \frac{2(\hat m+2 m_s)}{3(\hat m+m_s)} \tI^K
    \nn
     {\cal T}_{\gamma\gamma\rightarrow \pi^0 \eta}
      &=& -i \sqrt 3 e^2 k_1\cdot k_2 \frac{1}{f_\pi f_\eta} \tI^K
     \label{eq:one}
      \ee
with $k_1\cdot k_2 =\frac 12 (s-m_1^2-m_2^2)$. The one-loop 
finite contributions are
     \be
     \tI^i &\equiv & - H(s-4 m_i^2) \epsilon_1\cdot\epsilon_2 \frac{1}{16\pi^2}
     \left\{ 1+\frac{m_i^2}{s}
     \left(\ln\left[\frac{\sqrt s -\sqrt{s-4 m_i^2}}{\sqrt s +\sqrt{s-4 m_i^2}}
     \right] +i\pi\right)^2\right\} \nn
     && - H(4 m_i^2-s) \epsilon_1\cdot\epsilon_2 \frac{1}{16\pi^2}
     \left\{ 1-\frac{4 m_i^2}{s} 
      \arctan\sqrt{\frac{s}{4 m_i^2-s}}
     \right\} \ ,
     \label{eq:tI}
     \ee
where $H(x)$ is the Heaviside function. Following~\cite{YAZA} we used
the LHZ subtraction procedure. The ensuing counterterms (one) are fixed
by electric charge conservation. The results are independent of the
the parameters $C$ and $K$ introduced in (\ref{eq:sigma}). They are also
in agreement with one-loop chiral perturbation theory (ChPT)~\cite{TWO}
modulo counterterms. In ChPT all possible counterterms commensurate with 
symmetry and power counting are retained, in our case only those that 
show up in the loop expansion (minimal). Which of which is relevant is 
only determined by comparison with (threshold) experiments. Below, we will
show that both procedures yield almost identical results.

\section{Resonance Saturation Result}
\label{sec:5}

\begin{figure}[Th]
\centerline{\epsfig{file=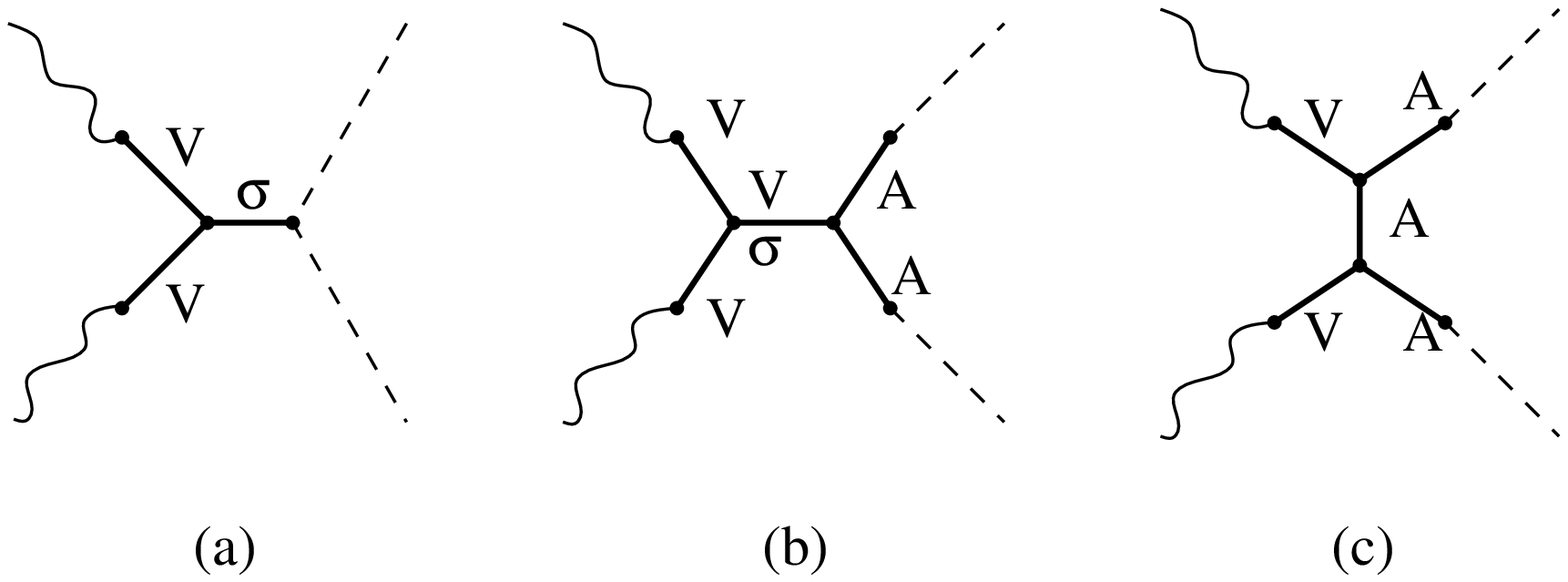,height=4cm}}
\vskip 5mm
\caption{Diagram for $\langle\V\V\sigma\rangle$ (a) and
 $\langle\V\V\j_A\j_A\rangle$ (b,c).  }
\label{fig:diag}
\end{figure}

To be able to address the fusion reaction processes beyond threshold
we need to take into account the  final state interactions in
(\ref{eq:T2}). One way to do this is to use dispersion analysis for
the three- and four-point functions with minimal weight-insertions.
This is equivalent to a tree-level resonance saturation of the three- 
and four-point functions as shown in Fig.~\ref{fig:diag} with all
possible crossings. Note that contact interactions are covered by
the present description in the limit where the masses of the
$\sigma$, $V$ and $A$ are taken to be very large.

{}From quantum numbers and parity, the vector current $\V^a_\mu$ 
will be saturated by the light vector mesons 
($v_\mu^a = \rho, \omega, \phi$), and the one-pion reduced
axial-vector current ${\j_A}^a_\mu$ by the light axial-vector mesons
($a_\mu^a = A_1, K_1$) \cite{LYZ98}. Typically,
     \be
     \langle 0 | \V_\mu^a (x) |v^b_\nu(p)\rangle
     &\sim& g_{\mu\nu} 
     \delta^{ab}\epsilon_\mu^V f_{v_a} m_{v_a} e^{-ip\cdot x} 
     \nn
     \langle 0 | {\j_A}_\mu^a (x) |a^b_\nu(p)\rangle
     &\sim& g_{\mu\nu}
     \delta^{ab}\epsilon_\mu^A f_{a_a} m_{a_a} e^{-ip\cdot x}\ .
     \ee
Since the photon carries indices  $c,d=3,8$ 
     \be
     v^3 &=& \rho^0 \nn
     v^8 &=& \sqrt{\frac 13}\omega^0 -\sqrt{\frac 23}\phi\ ,
     \ee
only the chargeless vector mesons contribute to the fusion process. 
The occurence of the structure constant $f^{abc}$ in the reduction of the
fusion reaction forces the axial-vector mesons to carry indices
$3,8$ as well. Hence, only $ a^{1,2} = A_1$ and $ a^{4\sim 7} = K_1$
will be needed in our case.

Similarly for $\V\V\j_A\j_A$ with
     \be
     \langle 0 | \V^d_\mu \V^c_\nu {\j_A}^b_\delta {\j_A}^a_\gamma |0\rangle 
     = \epsilon_\mu^V\epsilon_\nu^V \epsilon_\delta^A\epsilon_\gamma^A
     f_{a_a} f_{a_b} f_{v_c} f_{v_d}
     m_{a_a} m_{a_b} m_{v_c} m_{v_d}
     \langle 0| v^d_\mu v^c_\nu a^b_\delta a^a_\gamma |0\rangle\ .
     \ee

Finally, the scalar field $\hat \sigma$ can be saturated
by scalar mesons giving $\V\V\sigma$ as 
     \be
     \langle \V^d_\mu \V^c_\nu \hat \sigma\rangle 
     = \epsilon_\mu^V\epsilon_\nu^V 
      f_{v_c} f_{v_d} m_{v_c} m_{v_d}
     \langle v^d_\mu v^c_\nu \sigma \rangle\ .
     \ee
All the mesons will have masses and widths fixed at their PDG
(Particle Data Group) values.

With the above in mind the various contributions from Fig.~\ref{fig:diag}
can be readily constructed. In Appendix C we show how they could also be
retrieved using a linear sigma-model. The contribution to $\V\V\sigma$ is
     \be
     {\cal T}_{vv\sigma}^{ab} &=& 
      4 i e^2 \left( c_0  \delta^{ab}\delta^{cd} \delta^{h0} 
      + c_h d^{abh} d^{cdh} \right)
       \epsilon_1\cdot\epsilon_2  \frac{\Lambda M_b }{E_a E_b}
      \frac{f_{v_c} f_{v_d}}{m_{v_c} m_{v_d}}
      \frac{1}{s-m_{\sigma_h^2}} 
     \label{eq:r1} \ .
     \ee
The insertions of powers of $1/E$ and the scale 
$\Lambda$ ($= 1$ GeV) are to make the
arbitrary parameters $c_h$ (two) dimensionless. They will be fixed by
threshold constraints. The scalar contribution
to $\V\V\j_A\j_A$ is
     \be
     {\cal T}_{vvaa,\sigma_0}^{ab} &=& i 16 e^2
     \left( g_1 \delta^{cd}\delta^{ab}\delta^{h0} + g_2 d^{cdh} d^{abh}\right)
     \epsilon_1\cdot\epsilon_2
     \frac{\Lambda^2}{E_a E_b} 
     \frac{ f_{v_c} f_{v_d}}{m_{v_c} m_{v_d}}
     \frac{1}{s-m_{\sigma_h}^2}
     \nn &&
     \times
     k_1\cdot k_2\left(1-\frac{m_a^2}{m_{a_a}^2}-\frac{m_b^2}{m_{a_b}^2}
            +\frac{m_a^2}{m_{a_a}^2}\frac{m_b^2}{m_{a_b}^2}\right)
     \frac{f_{a_a}m_{a_a}}{m_a^2-m_{a_a}^2} 
     \frac{f_{a_b}m_{a_b}}{m_b^2-m_{a_b}^2} \ .
     \label{eq:r2}
     \ee
The dimensionless parameters $g_h$ (two) are again arbitrary.
The intermediate vector contribution from Fig.~\ref{fig:diag}-(b) 
vanishes because of the antysymmetry of the structure constant 
$f^{abc}$. Finally, the contribution from Fig.~\ref{fig:diag}-(c) is
     \be
     {\cal T}_{vvaa,a}^{ab} &=& -i 16 e^2 g_3 f^{caf} f^{dbf}
     \frac{1}{E_a E_b}
     \left(\frac{f_{v_c} f_{v_d}}{m_{v_c} m_{v_d}}\right)
     \left(\frac{1}{t-m_{a_f}^2}\right)
     \left(\frac{ f_{a_a} m_{a_a}}{m_a^2-m_{a_a}^2}\right) 
     \left(\frac{ f_{a_b} m_{a_b}}{m_b^2-m_{a_b}^2}\right) 
     \nn
     && \times \left[
       (\epsilon_1\cdot k_1) (\epsilon_2\cdot k_2)
          \left(1-\frac{t}{m_{a_f}^2}\right)
      \left( -t + m_a^2 +m_b^2
          -\frac{(k_1\cdot q_1)^2}{m_{a_a}^2}
          -\frac{(k_2\cdot q_2)^2}{m_{a_b}^2} \right) \right.
     \nn && \left.
      +\left(\epsilon_1\cdot\epsilon_2
       +\frac{ (\epsilon_1\cdot k_1) (\epsilon_2\cdot k_2)}{m_{a_f}^2}\right)
       \left( m_a^2-\frac{(k_1\cdot q_1)^2}{m_{a_a}^2}\right)
       \left( m_b^2-\frac{(k_2\cdot q_2)^2}{m_{a_b}^2}\right)
      \right]
     \nn &&
     + (t,a,k_1\leftrightarrow u,b,k_2) 
     \label{eq:r3}
     \ee
with one additional dimensionless parameter $g_3$.

In the vector and axial channels all the resonances quoted above are
introduced with their masses and decay witdths in the form of Breit-Wigner
resonances fixed at their PDG values. 
In the scalar channels we will use three resonances for 
$\sigma^0$ : $f_0 (500)$, $f_0 (980)$ and $f_2 (1270)$. As our chief goal
is to test the master formula result with resonance saturation, we will 
keep our description simple by substituting
    \be  
    \frac{1}{s-m_{\sigma_0}^2} \rightarrow
    \sum_{m_f}  \frac{f_f}{s-m_{f}^2 +i G (s,m_f) m_{f}}
    \ee
with $f_{f_0(500)}=f_{f_2(1270)}=1$ and $f_{f_0(980)}=0.05$, and 
the decay widths 
    \be
    G(s,m_f) = H(s-4 m_\pi^2) G_0 
    \left(\frac{1- 4 m_\pi^2/s}{1-4 m_\pi^2/m_f^2}\right)^n \ ,
    \ee
with $n=1/2$ and $3/2$ for scalar and vector mesons, respectively. 
A more detailed parametrization of the partial widths and so on
will not be attempted here, again for simplicity.
We have found that the contribution of $f_0(980)$ is suppressed
(hence the order of magnitude change in the weight) in agreement
with previous investigations~\cite{OSET}.
In the numerical analysis to follow, we have checked that our
results are not greatly sensitive to the resonance parametrizations
provided that PDG masses and widths are enforced.

In the isotriplet-scalar channel $\sigma^3$ we have : $a_0(980)$ and 
$a_2(1320)$, giving
    \be  
    \frac{1}{s-m_{\sigma_3}^2} \rightarrow
    \frac{0.6}{s-m_{a_0}^2 +i G_{a_0} (s) m_{a_0}}
    -\frac{1}{s-m_{a_2}^2 +i G_{a_2} (s) m_{a_2}} \ .
    \label{isosign}
    \ee
The same functional form for $G$ is used, but with a different cut-off 
corresponding to the lowest mass yields in the various decay channels.
The relative sign in (\ref{isosign}) reflects the attractive character
of $a_2(1320)$ in comparison to $a_0(980)$ in the isotriplet channel.
We will not consider the effects of $\sigma^8$ as it involves higher
octet-scalar resonances.

\section{Polarizabilities}
\label{sec:6}

Before discussing in details how our analysis of the fusion reactions
compare in details to the present data, we will first address the issue
of the meson polarizabilities as inferred from our one-loop analysis.
For the charged pions~\cite{dono93},
   \be
   \bar\alpha_E^{\pi^\pm} &=& (6.8\pm1.4\pm1.2)\times 10^{-4}\; {\rm fm}^3
   \nonumber\\
   \bar\alpha_E^{\pi^\pm} &=& (20\pm 12)\times 10^{-4}\; {\rm fm}^3
   \nonumber\\
   \bar\alpha_E^{\pi^\pm} &=& (2.2 \pm 1.6)\times 10^{-4}\; {\rm fm}^3 \ ,
   \ee
and for the neutral pions~\cite{dono93}
   \be
   |\bar\alpha_E^{\pi^0}| &=& (0.69\pm 0.07\pm 0.04)\times 10^{-4}\; {\rm fm}^3
   \nonumber\\
   |\bar\alpha_E^{\pi^0}| &=& (0.8\pm 2.0)\times 10^{-4}\; {\rm fm}^3 \ .
   \ee
The data are not accurate enough. This notwithstanding, our one-loop 
result for the charged pions is
   \be
   \alpha_L^{\pi^\pm} \approx  4.2\times 10^{-4}\; \fm^3 \ ,
   \ee
this is twice the value obtained using standard chiral perturbation theory
 \cite{dono93}. The
difference stems for the additional (finite) counterterms in ChPT, which are
purposely absent (minimal) in our analysis. This point was discussed in great
details in \cite{YAZA}. For the neutral pions we have
   \be
   \alpha_L^{\pi^0} \approx  6.3\times 10^{-4}\; \fm^3 \ .
   \ee
For the rest of the octet, we have
   \be
   \alpha_L^{K^\pm} &\approx & - 2.7 \times 10^{-5}\, \fm^3 \nonumber\\
   \alpha_L^{K^0 \bar K^0} &\approx & + 2.8 \times 10^{-5}\, \fm^3 \nonumber\\
   \alpha_L^{\eta} &\approx & - 4.4 \times 10^{-6}\, \fm^3 .
   \ee
In the resonance saturation approach, the polarizabilities follow
essentially from the ${\bf V V}{\bf j}_A{\bf j}_A$ contributions
in (\ref{eq:r3}). These contributions are constrained at high energy
to be small, resulting into naturally small polarizabilities. A global
fit yields pion polarizabilities that are similar for charged and 
chargeless fusion reactions.

\section{Numerical Results}
\label{sec:7}

Most of the calculations to be discussed in this section are carried
with the PDG parameters for the quoted resonances. The dimensionless
couplings involved in the resonance saturation approach are chosen
so as to give a global fit that is consistent with the threshold
constraints (mainly one-loop). Specifically, we will use
   \be
   c_0 &=& -98208 \nonumber \\
   c_3 &=& 5 c_0 \nonumber \\
   g_1 &=& 0.9744 \nonumber\\
   g_2 &=& -13.64 \nonumber\\
   g_3 &=& -1.5 \ ,
   \ee 
with $c_8=0$ since we are ignoring the effects from $\sigma^8$. Some of the 
results for the total cross sections to be quoted will involve a parameter $Z$ 
defined as
\be
\sigma_Z = 2\int_0^Z d{\rm cos}\theta\, \frac{d\sigma}{d{\rm cos}\theta}
\ee
where $\theta$ is the relative angle between one of the two incoming photons 
and the outgoing mesons.

\subsection{Pions}

\begin{figure}
\begin{center}
            \epsfig{file=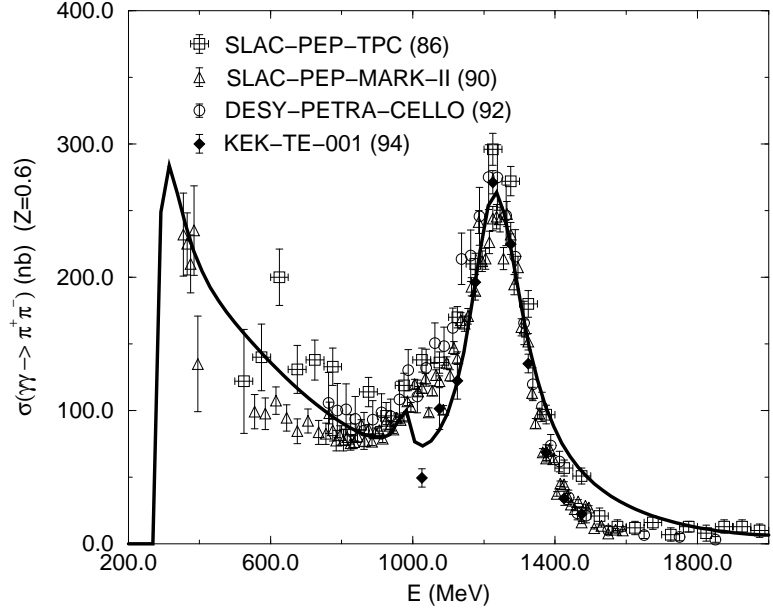,width=4.5in}
            \epsfig{file=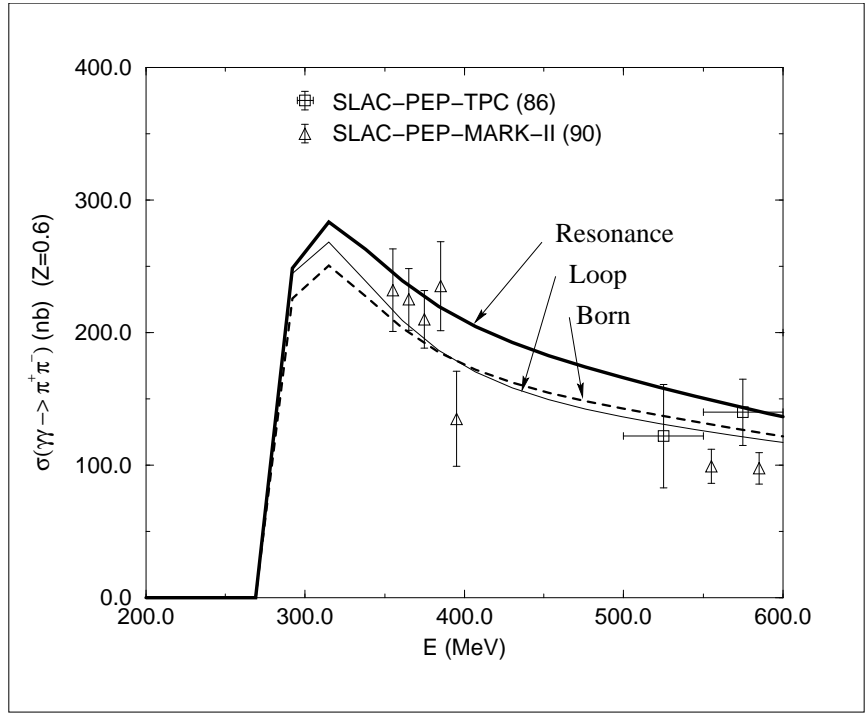,width=4.5in}
\end{center}
\caption{
 Total cross section for $\gamma\gamma\rightarrow\pi^+\pi^-$ (Z = 0.6).
 Thick (thin) line correspond to resonance (loop) contribution.
 Dashed line in the lower panel corresponds to the Born term.
 The data are collected from Refs.~\protect\cite{TPC,MARK,DESY,KEK}
}
\label{fig:spipm}
\end{figure}

In Fig.~\ref{fig:spipm} we show the total cross section for fusion to 
charged pions
up to $Z=0.6$. The data are from ~\cite{TPC,MARK,DESY,KEK}. The 
overall agreement with the data is good. Our analysis appears to favor the 
SLAC-PEP-MARK-II  as well as the KEK-TE-001 data. The peak at $f_2 (1270)$
is clearly visible, while the $f_0 (980)$ is weaker. The Born contribution
overwhelms the $f_0 (500)$ contribution in this channel, and is hardly
visible in our results as well as the data. In the insert, we show an 
enlargement of the threshold region and comparison with our Born contribution, 
the one-loop analysis, and the resonance saturation approach. Overall, our 
approximations are consistent.

\begin{figure}
\begin{center}
            \epsfig{file=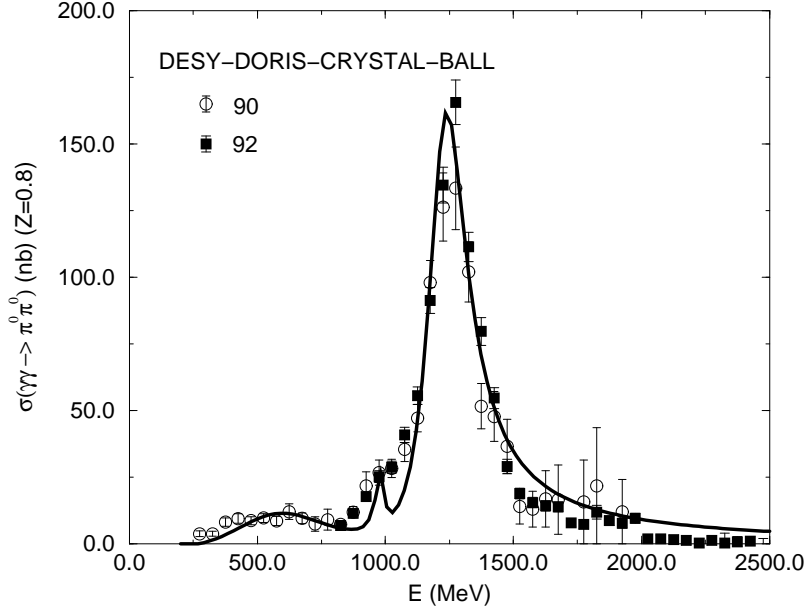,width=4.5in}
            \epsfig{file=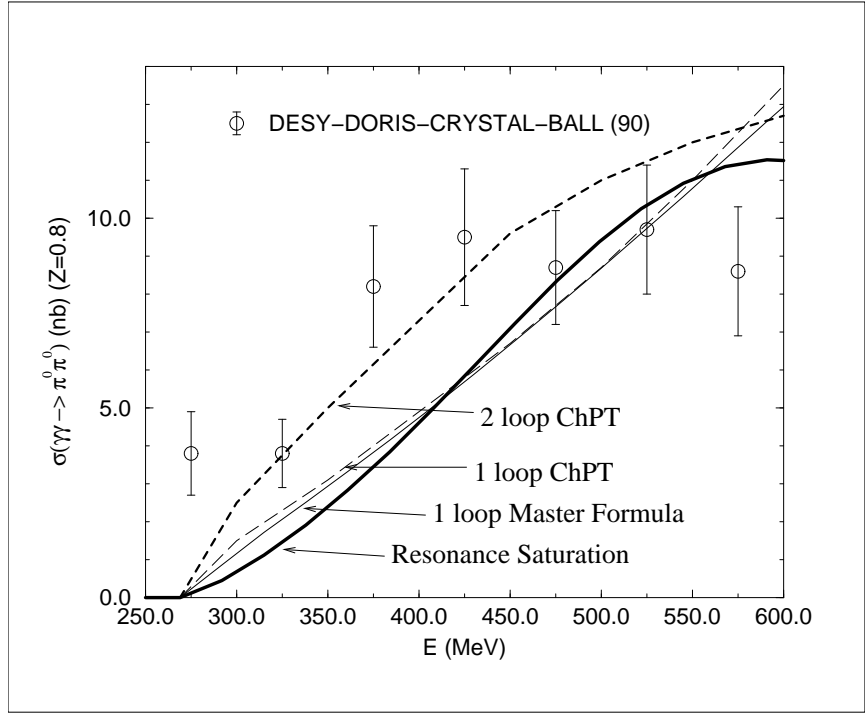,width=4.5in}
\end{center}
\caption{
 Total cross section for $\gamma\gamma\rightarrow\pi^0\pi^0$ (Z = 0.8).
 Thick (thin) line correspond to resonance (loop) contribution.
 The dashed lines in the lower panel are the 1- and 2-loop ChPT
 results \protect\cite{TWO}.
 The data are taken from Refs.~\protect\cite{DORIS,SAN}.
}
\label{fig:spizz}
\end{figure}

In Fig.~\ref{fig:spizz} we present our results for the fusion reaction into
chargeless pions with $Z=0.8$. The data are from~\cite{DORIS,SAN}. Again the
$f_2 (1270)$ is clearly visible, while the $f_0 (980)$ is barely.
The broad effects from the $f_0 (500)$ are also visible in comparison to the 
data. The resonance saturation result is in overall agreement with the both
sets of data. In the insert, we show an enlargement of the threshold region
and comparison to our one-loop result as well as one-loop and two-loop ChPT.
Clearly our one-loop and the one-loop ChPT are in good agreement although our
construction is minimal (fewer counterterms). Most of the parameters (six)
in the two-loop results from ChPT are fit using ideas similar to the resonance
saturation approach we have adopted.

\subsection{Kaons}

\begin{figure}
\begin{center}
\epsfig{file=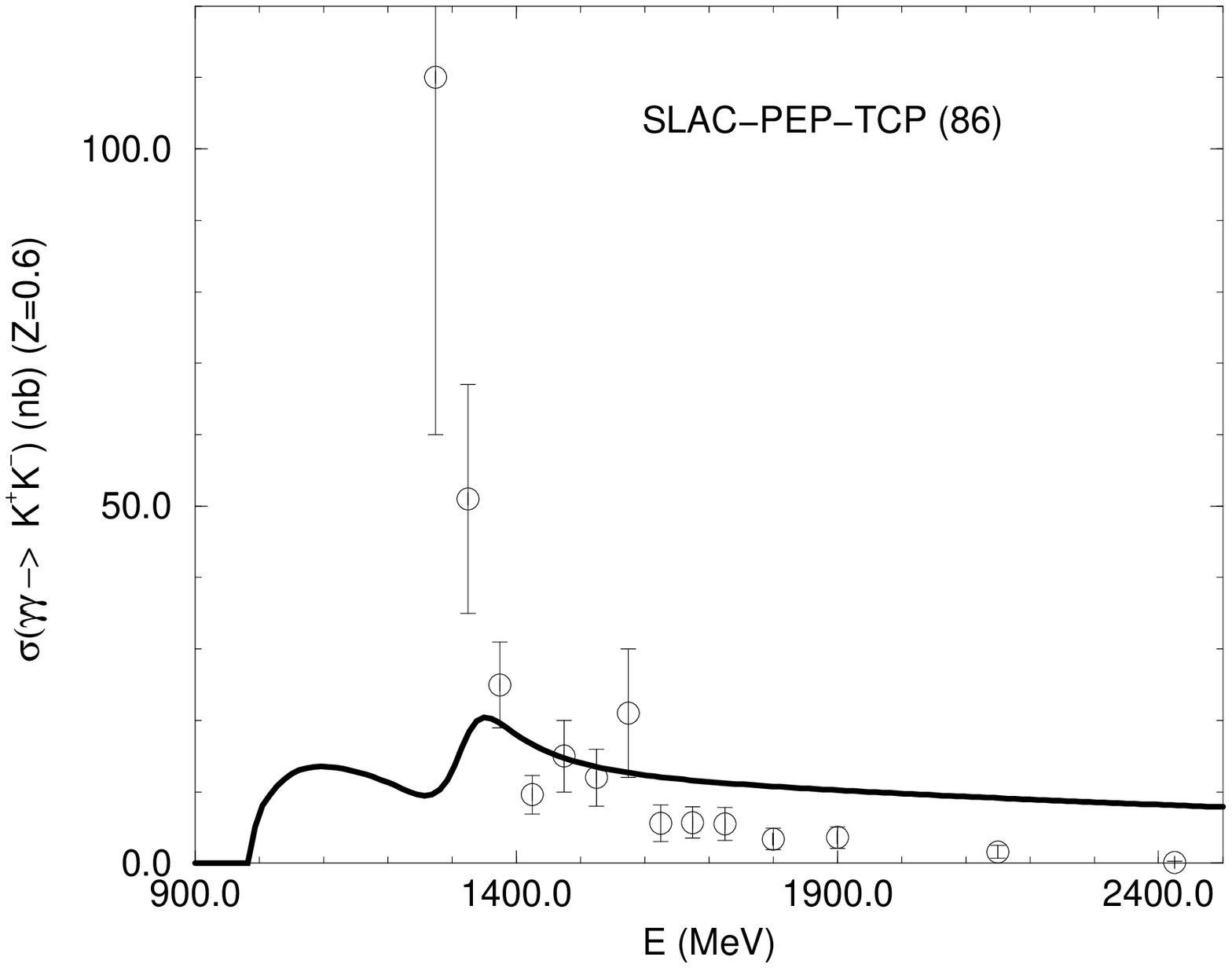,width=4.5in}
\epsfig{file=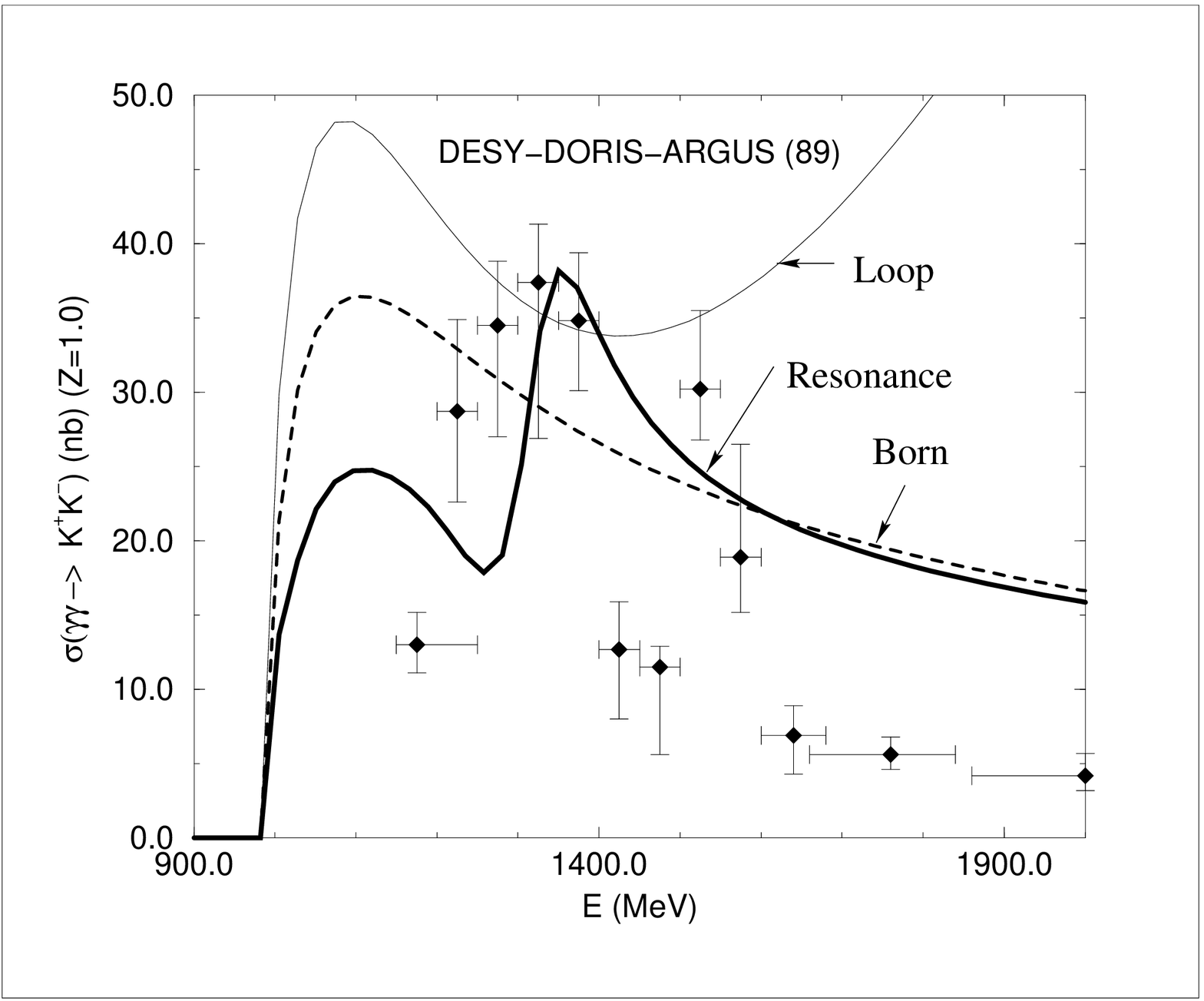,width=4.5in} 
\end{center}
\caption{Total cross section for 
$\gamma\gamma\rightarrow K^+K^-$  with Z=0.6 and Z=1.0. 
 Thick (thin) line corresponds to resonance (loop) contribution.
 The Born terms are plotted as dashed line.
 The data are taken from Refs.~\protect\cite{TPC,KKpm1}.
}
\label{fig:sKpm}
\end{figure}

In Fig.~\ref{fig:sKpm}  we present our results for the fusion reaction into 
charged kaons. For $Z=0.6$ our analysis shows a treshold enhancement at about
$980$ MeV, followed by another enhancement at $a_2(1320)$. The enhancement
shown in the SLAC-PEP-TCP data is consistent with the $a_2(1320)$, although
the error bars are large. Our results
for the cross section are higher than the data in the energy range 
$\sqrt{s}=1.6-2.4$ GeV. For $Z=1.0$ we compare the resonance saturation results 
with the Born amplitude and the one-loop approximation. Again we see the same 
features as those encountered at $Z=0.6$. The DESY-DORIS-ARGUS data agree with
our analysis around the $a_2(1320)$, but are not in agreement at threshold and
above $\sqrt{s}=1.6$ GeV. The threshold enhancement due to the Born terms in 
our analysis is only partly decreased by the repulsive character of the 
scalar-isotriplet $a_0(980)$. A similar behaviour was also noted by Oller
and Oset~\cite{OSET} using a coupled channel analysis.

\begin{figure}
\begin{center}
\epsfig{file=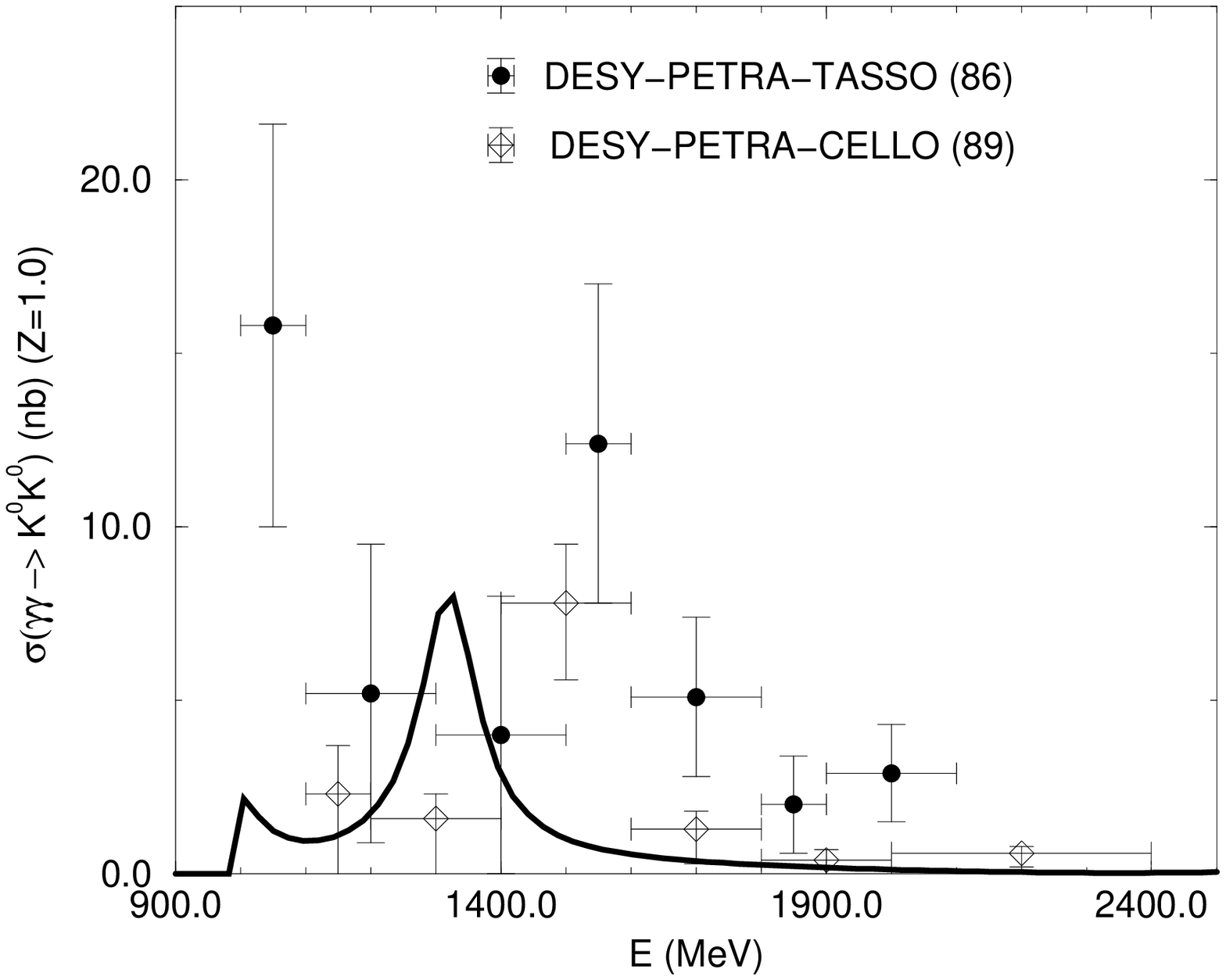,width=4.5in}
\end{center}
\caption{Total cross section for 
$\gamma\gamma\rightarrow K^0\bar K^0$ (Z=1.0).
 The data are taken from Refs.~\protect\cite{TASSO,CELLO}.
}
\label{fig:K0K0}
\end{figure}

In Fig.~\ref{fig:K0K0} we show our results for chargeless kaons. Our results
favor the data from DESY-PETRA-CELLO~\cite{CELLO} as opposed to the early
data from DESY-PETRA-TASSO~\cite{TASSO}, although the data have large error
bars. The effects from the $a_0(980)$ is weaker than the one from the 
$a_2(1320)$. In this case the Born contribution vanishes.

\subsection{Etas}

\begin{figure}
\begin{center}
\epsfig{file=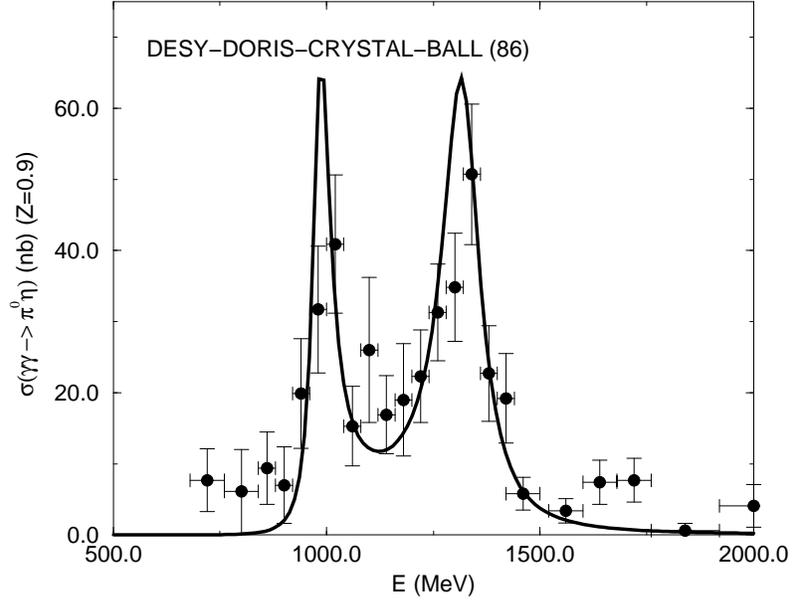,width=4.5in}
\end{center}
\caption{Total cross section for 
$\gamma\gamma\rightarrow \pi^0\eta$ (Z=0.9).
The data are taken from Refs.~\protect\cite{pieta}.
}
\label{fig:spieta}
\end{figure}

\begin{figure}
\begin{center}
\epsfig{file=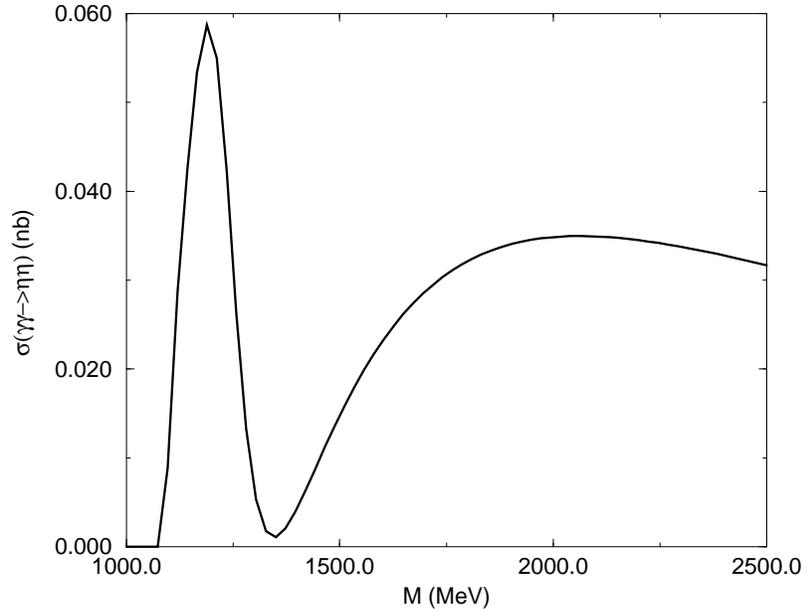,width=4.5in}
\end{center}
\caption{Total cross section for 
$\gamma\gamma\rightarrow \eta\eta$.}
\label{fig:setae}
\end{figure}

In Fig.~\ref{fig:spieta} we show our results for the fusion into $\pi^0\eta$
for $Z=0.9$. The peaks are the scalar-isotriplets $a_0(980)$ and $a_2(1320)$.
There is fair agreement with the DESY-DORIS-CRYSTAL-BALL~\cite{pieta} data.
The strength between the two-resonances follow simply from the relative
sign in (\ref{isosign}) reflecting on the attraction-repulsion in these 
two channels. In Fig.~\ref{fig:setae} we show our predictions for the fusion 
reaction to two eta's. The cross section is tiny in comparison to the other
fusion reactions (about four orders of magnitude down). 
The reason is the near cancellation between the $f_2 (1270)$
contribution in ${\bf V}{\bf V}\sigma$ and ${\bf V}{\bf V}{\bf j}_A{\bf j}_A$
($c_0$ and $g_1$ have opposite signs). Since the resonance is smeared 
differently in the two contributions, the exact cancellation takes place
in the range $1.25-1.5$ GeV.

\section{Conclusions}
\label{sec:8}

We have analyzed the two-photon fusion reaction to two mesons using the
master formulae approach to QCD with three flavors. The formulae for the
fusion reaction amplitude encodes all the information about chiral symmetry
and its breaking in QCD. We have analyzed this result in power counting and 
shown that it is overall in agreement with results from three-flavor ChPT
in the threshold region. We have derived specific results for the real part
of the polarizabilities of all the octet mesons.

To analyze the reactions beyond threshold, we have implemented a simple 
dispersion analysis on the pertinent three- and four-point functions in the
form of tree-level resonance saturation. The analysis enforces broken chiral
symmetry, unitarity and crossing symmetry in a staightforward way. The 
pertinent resonance parameters (masses and widths) are fixed at their PDG
values. Their couplings result into 5 parameters which we use to globally 
fit all available data through $\sqrt{s}= 2$ GeV and predict a very small 
cross section for $\gamma\gamma\rightarrow \eta\eta$. 

The master formulae to the fusion reaction processes implies from first 
principles scalar-isoscalar and scalar-isotriplet correlations in the
s-channel, and axial-vector correlations in t-channels. The latters
enforce the correct polarizabilities, while the formers account for most
of the resonances seen in the experiments. In particular, the scalar-isoscalar
$f_0 (500)$, $f_0(980)$ and $f_2(1270)$ are predominant in the fusion
reactions involving pions, while the scalar-isotriplet $a_0(980)$ and 
$a_2(1320)$ are important in the fusion reactions involving kaons, and
also etas and pions. The $a_0(980)$ is found to decrease considerably the 
threshold enhancements caused by the Born term in the fusion to charged kaons,
in agreement with present experiments. The present results are important in
the assessment of the electromagnetic emission rates from a hadronic gas in 
relativistic heavy-ion collisions \cite{LYZ98X}.

\section*{Acknowledgement}
IZ would like to thank Jose Oller and Eulogio Oset for discussions.
This work was supported by the U.S. Department of Energy under Grant No.
DE--FG02--88ER40388.


\newpage

\setcounter{equation}{0}
\renewcommand{\theequation}{A.\arabic{equation}}
\section{Appendix A :  Details of the Born Contributions}

In this Appendix we detail the Born contributions to the T-matrix for the fusion 
process as given by (\ref{eq:T1}-\ref{eq:T2}). If we recall that the meson 
indices are $(a,b)$ and the photon indices $(c,d)$, then the contact 
contributions in (\ref{eq:T1}) are
    \be
    {\cal T}_{11} &= & 
       i g_{\mu\nu} (E)^a (E^{-1})^b (f^{bci} f^{ida} +f^{bdi} f^{ica} ) \nn
    &=& i2 \epsilon_1\cdot\epsilon_2 (E)^a (E^{-1})^b 
    \left(f^{b3i} +\frac{1}{\sqrt 3} f^{b8i}\right)
    \left(f^{ai3} +\frac{1}{\sqrt 3} f^{ai8}\right) \nn
    &=& - i 2 \epsilon_1\cdot\epsilon_2
    \times \left( \begin{array}{ccc}
                    \pi^\pm  &:& 1 \\
                    \pi^0    &:& 0 \\
                    K^\pm    &:& 1 \\
                    \bar K^0 &:& 0 \\
                    \eta     &:& 0 \; 
                    \end{array}
                    \right) \,,
    \ee
while the pole terms are
     \be
     {\cal T}_{12} &=& i (2 k_2-q_1)^\mu (2k_1-q_2)^\nu (E)^a (E^{-1})^b
     \left\{  f^{bci}f^{ida} \frac{1}{(k_2-q_1)^2-m_i^2} 
     +  f^{bdi}f^{ica} \frac{1}{(k_2-q_2)^2-m_i^2} \right\} \nn
     &= &
      4i \epsilon_1\cdot k_1\epsilon_2\cdot k_2 (E)^a (E^{-1})^b
     \left(f^{b3i} +\frac{1}{\sqrt 3} f^{b8i}\right)
     \left(f^{ai3} +\frac{1}{\sqrt 3} f^{ai8}\right)
     \left(\frac{1}{u-m_i^2}+\frac{1}{t-m_i^2}\right) \nn
     &=& -4i \epsilon_1\cdot k_1\epsilon_2\cdot k_2 
     \times \left(
     \begin{array}{ccc}
     \pi^\pm  &:& \left[(t-m_\pi^2)^{-1}+(u-m_\pi^2)^{-1}\right] \\
     \pi^0    &:& 0 \\
     K^\pm    &:& \left[(t-m_K^2)^{-1}+(u-m_K^2)^{-1}\right] \\
     \bar K^0 &:& 0 \\
     \eta     &:& 0
     \end{array}
     \right) 
     \ee
with $q\cdot\epsilon=0$. Only the charged contributions appear in the Born 
approximation, since photons do not couple to chargeless particles at tree
level. We note that the contact term in (\ref{eq:T2}) contributes
     \be
     {\cal T}_{21} &=& i g_{\mu\nu} \frac{1}{E^a E^b} f^{bdi}f^{aci}
       \left\{\frac 23 K \left(\frac{M}{m_p^2}\right)^a - E_i^2 \right\} \nn 
     &=& - i2 \epsilon_1\cdot\epsilon_2 \frac{1}{E^a E^b}
       \left\{\frac 23 K \left(\frac{M}{m_p^2}\right)^a - E_i^2 \right\}
       \left(f^{b3i} +\frac{1}{\sqrt 3} f^{b8i}\right)
       \left(f^{ai3} +\frac{1}{\sqrt 3} f^{ai8}\right) \nn
     &=& -i2 \epsilon_1\cdot\epsilon_2 
       \left\{1-\frac 23 K \left(\frac{M}{E^2 m_p^2}
       \right)^{\pi^\pm,K^\pm}  \right\} \ .
     \ee
In the SU(2) case we have 
$2K/3 \rightarrow ({f_\pi^2 m_\pi^2})/{\hat m}$,
and this extra contribution vanishes. In the SU(3) case
this extra term is of order $1/E^2$ in power counting. It renormalizes 
to zero when combined with the counterterms in 
$\langle\V\V\hat\sigma\rangle$ and $\langle\V\V\j_A\j_A\rangle$,
leaving the octet charges integer-valued.

\setcounter{equation}{0}
\renewcommand{\theequation}{B.\arabic{equation}}
\section*{Appendix B : Details of the One loop Contributions}

In this Appendix we give some details regarding the one-loop 
analysis carried in section IV, following the discussion in~\cite{YAZA}. 
In particular, we will assess the one-loop contribution to the three-
and four-point correlators $\langle \V\V\sigma\rangle$ and
$\langle\V\V\j_A\j_A\rangle$.

\subsection*{B-i. ${\bf V}{\bf V}\sigma$}

The one-loop contribution to $\V\V\sigma^{0,3,8}$ is generically of the form
     \be
     \langle 0 | T^\star \V_\mu^c(x) \V_\nu^d(y)\hat\sigma_h (z) 
                                                       |0\rangle_{\rm con.}
      &=& \left[\frac{\delta \K^{ii}}{\delta Y^h}\right]
         f^{ijc}f^{ijd}\left(
         g_{\mu\nu}\delta^4(x-y)
         \int_q e^{-iq\cdot (y-z)} {\cal I}^{ii} (q) \right. \nn
     &&+\left. \int_p \int_q e^{iq\cdot x -iq\cdot y -i (q-p)\cdot z} 
         {\cal I}_{\mu\nu}^{ij} (q,p)\right) \ ,
     \ee
with $(c,d)=(3,8)$, $\int_q \equiv \int {d^4 q}/{(2\pi)^4}$,
and $\K$ is the SU(3) version of (3.39) in Ref.~\cite{YAZA}.
Specifically,
     \be
     \K^{ac} &=& 2 \uv_\mu^{ac}\partial^\mu +\partial^\mu\uv_\mu^{ac}
     +\uv^{\mu ab}\uv_\mu^{bc}-\ua^{\mu ab}\ua_\mu^{bc}
     -\frac{C}{K}Y^0\left(\frac{m_p^2}{M}\right)^{ac}
     +Y^b \hat d^{abd} (E^{-1})^{dc} ,
     \ee
with $\underline{A}_\mu^{ac} = A_\mu^b f^{abc}$ and
     \be
     \hat d^{abc} &=& d^{abc} \left(\frac{M}{E m_p^2}\right)^b
         \left(\frac{E m_p^2}{M} \right)^c .
     \ee
Here the "$\sigma^3$" contribution vanishes  because the
$K^\pm$ and $K^{0,\bar0}$ contribution cancel within the loop. 

The one-loop integrals are
     \be
     {\cal I}^{ij} (q) &=& -i \int_k \left( \frac{1}{k^2-m_j^2+i0} \cdot
               \frac{1}{(k+q)^2-m_i^2+i0} - (q=0)\right)
     \nn
     {\cal I}_{\mu\nu}^{ij} (q,p) 
         &=& i \int_k \left( \frac{1}{k^2-m_j^2+i0} \cdot
            \frac{(2k+q)_\mu}{(k+q)^2-m_i^2+i0} \cdot
            \frac{(2k+p)_\nu}{(k+p)^2-m_i^2+i0} - (q=p=0)\right).
     \ee
Each integral is made finite by one subtraction at $q=0$ (first) and
$q=p=0$ (second). This results in one counterm which renormalizes the
charge to its integer value. In the LHZ scheme followed
here charge conservation is not protected by logarithmic
divergences. 

The final expressions in (\ref{eq:tI}) are quoted in terms of 
     \be
     \tI^i &\equiv & \epsilon_1\cdot\epsilon_2 {\cal I}^{ii} (q_1+q_2) 
         +\epsilon_1^\mu \epsilon_2^\nu {\cal I}_{\mu\nu}^{ii} (q_1,-q_2) \ .
     \ee
Also the contributions from the three-point function in (\ref{eq:T2}) read
     \be
     {\cal T}_4^{\gamma\gamma\pi^0\pi^0} 
     &=& -2 i \frac{m_\pi^2}{f_\pi^2} \tI^\pi
         - i \frac{m_K^2}{f_\pi^2} \frac{2\hat m}{\hat m+m_s}\tI^K \nn
         {\cal T}_4^{\gamma\gamma\pi^+\pi^-} 
     &=& {\cal T}_4^{\gamma\gamma\pi^0\pi^0} \nn
         {\cal T}_4^{\gamma\gamma K K} &=& 
         -i \frac{m_\pi^2}{f_K^2}\frac{\hat m+m_s}{2\hat m}\tI^\pi 
         -i \frac 3 2 \frac{m_K^2}{f_K^2}\tI^K \nn
     {\cal T}_4^{\gamma\gamma \eta \eta} 
     &=& -i\frac 23 \frac{m_\pi^2}{f_\eta^2}
         \frac{\hat m+2 m_s}{3\hat m}\tI^\pi
         -i\frac{5}{3}\frac{m_K^2}{f_\eta^2}
         \frac{2(\hat m+2 m_s)}{3(\hat m+m_s)}\tI^K \nn
     {\cal T}_4^{\gamma\gamma \pi \eta} 
     &=& 0 \ .
     \ee

\subsection*{B-ii. ${\bf V}{\bf V}{\bf j}_A{\bf j}_A$}

The one-loop contribution to the four-point function
$\V\V\j_A\j_A$ maybe obtained similarly. In particular,
     \be
     \langle \V\V\j_A\j_A \rangle
     &\equiv & \langle 0 | T^\star \V_\mu^a(x) \V_\nu^b(y) 
           {\j_A}_\alpha^c (z_1) {\j_A}_\beta^d (z_2) |0\rangle_{\rm con.} \nn
     &=& -i (f^{iah}f^{hbj}+f^{ibh}f^{haj})\cdot
      (f^{jcl}f^{ldi}+f^{jdl}f^{lci})\delta^4(z_1-z_2) g_{\alpha\beta} \nn
     && \times   \left( g_{\mu\nu}\delta^4(x-y)
       \int_q e^{-iq\cdot (y-z)} {\cal I}^{ij} (q)
       +\int_p \int_q e^{iq\cdot x -iq\cdot y -i (q-p)\cdot z} 
        {\cal I}_{\mu\nu}^{ij} (q,p) \right) \ .
    \ee
Since we need the integrated version, then
    \be
    i \int \langle \V\V\j_A\j_A\rangle
    &\equiv& 
    i \int_x \int_y e^{iq_1\cdot x-i q_2\cdot y} 
      \left\langle 0\left|\hS \T^\star \left[
      \left(\V^{\mu,3}(x)+\frac{1}{\sqrt 3}\V^{\mu,8}(x) \right)
    \right.\right.\right. \nn
    && \times \left.\left.\left.
      \left(\V_\nu^3(y)+\frac{1}{\sqrt 3}\V_\nu^8(y) \right)
      {\j_A}_\alpha^c(z){\j_A}_\beta^d(0)\right] \right| 0\right\rangle
    \nn
    &=& 2 \times \left\{
    \begin{array}{|c|c|c|}
    \hline
     (c,d) & {\rm SU(3) meson} & {\rm contribution}  \\
    \hline
     (1,1),(2,2) & \pi^\pm & {\cal I}^\pi + \frac 12 {\cal I}^K \\ 
     (3,3) & \pi^0   & 2 {\cal I}^\pi + \frac 12 {\cal I}^K \\ 
     (4,4),(5,5) & K^\pm   & \frac 12 {\cal I}^\pi + {\cal I}^K \\ 
     (6,6),(7,7) & K^{0\bar 0} & \frac 12 {\cal I}^\pi + \frac 12{\cal I}^K \\ 
     (8,8) & \eta & \frac 3 2 {\cal I}^K \\
     (3,8) & (\pi^0,\eta) & \frac{\sqrt 3}{2} {\cal I}^K \\
    \hline
    \end{array} \right.
    \ee
where ${\cal I}^i$ ($\tilde {\cal I}^i (q_1,q_2) \equiv {\cal I}^i(q_1,-q_2)$)
is defined as
    \be
    {\cal I}^{i} (q_1,q_2)
    &\equiv & g_{\mu\nu} {\cal I}^{ii} (q_1-q_2) 
              + {\cal I}_{\mu\nu}^{ii} (q_1,q_2)\ .
    \ee
The respective contributions to (\ref{eq:T2}) 
from $\langle \V\V\j_A\j_A\rangle$ are
    \be
    \begin{array}{ccl}
(\pi^0,\pi^0)  & : & -i2 k_1\cdot k_2 f_\pi^{-2} (2\tI^\pi + \frac 12 \tI^K) \\
(\pi^+,\pi^-)  & : & -i2 k_1\cdot k_2 f_\pi^{-2} (\tI^\pi +  \frac 12 \tI^K) \\
(\K^+,\K^-)    & : & -i2 k_1\cdot k_2 f_K^{-2} (\frac 12 \tI^\pi+ \tI^K) \\
(K^0,\bar K^0) & : & -i2 k_1\cdot k_2 f_K^{-2} 
                         (\frac 12 \tI^\pi+\frac 12 \tI^K) \\
(\eta,\eta)    & : & -i2 k_1\cdot k_2 f_\eta^{-2} ( \frac 3 2\tI^K) \\
(\pi^0,\eta)   & : & -i2 k_1\cdot k_2 f_\pi^{-1} f_\eta^{-1}
                         ( \frac{\sqrt 3}{2}\tI^K )\\
    \end{array}
    \ee
with $k_1\cdot k_2 =  (s- m_1^2-m_2^2)/2$.

\setcounter{equation}{0}
\renewcommand{\theequation}{C.\arabic{equation}}
\section*{Appendix C : $\Sigma$ model}

In this Appendix we provide a simple implementation
of the resonance saturation analysis at tree level in
the context of the linear sigma-model. This complements
our general analysis in section V.

\subsection*{C-i. Lagrangian}

Consider the linear sigma-model with general (quadratic) couplings 
to vector and axial-vectors with global chiral symmetry
    \be
    {\cal L}_{\Sigma} 
      &=& \frac 14 \Tr\left[D_\mu \Sigma D^\mu\Sigma^\dagger\right] \nn
   {\cal L}_{kin} &=&
     \Tr \left(-\frac 14 (F_{l,r}^{\mu\nu})^2 
     +\frac{m^2}{2} (A_{l,r}^\mu)^2\right) \nn 
   {\cal L}_{int} &=&
      \frac 14 b_1 g^2 \Tr\left[\Sigma\Sigma^\dagger\right]
       \Tr\left( (A_{l,r}^\mu)^2\right)
      -b_2 g^2 \Tr\left[A_l^\mu\Sigma A_r^\mu\Sigma^\dagger\right] 
   \nn &&
     -b_3 g^2 \Tr\left[A_l^\mu A_{l,\mu} \Sigma\Sigma^\dagger\right]
     -b_4 g^2 \Tr\left[A_r^\mu A_{r,\mu} \Sigma^\dagger\Sigma\right]
   \ee
where 
   \be
   \Sigma &=& (\hat \sigma^0 -C) \one +\sigma^h\lambda^h +i\pi^a\lambda^a
   \nn
   D^\mu\Sigma &=& \partial^\mu \Sigma-ig
     (A_l^\mu\Sigma-\Sigma A_r^\mu)
   \nn
   F_{l,r}^{\mu\nu} &=& \partial^\mu A_{l,r}^\nu
    -\partial^\nu A_{l,r}^\mu -ig \left[A_{l,r}^\mu,A_{l,r}^\nu\right],
   \ee
where $-C$ is the vacuum expectation value of $\sigma^0$, and
the vector field is given as
   \be
   A_{l,r}^\mu = (\vec v^\mu \pm \vec a^\mu)\cdot \vec\lambda \ .
   \ee

The $vv\sigma$-vertex comes only from ${\cal L}_{int}$,
   \be
   {\cal L}_{vv\sigma} &=& 
    4 (b_1-b_2-b_3-b_4) g^2 C v_\mu^a v^{\mu,a}\hat\sigma_0
    -  4 (b_2 +b_3+b_4  ) g^2 C d^{abh} v_\mu^a v^{\mu,b} \sigma^h \ .
   \ee
The $aa\sigma$-vertex comes from both ${\cal L}_{\Sigma}$ and ${\cal L}_{int}$,
   \be
   {\cal L}_{aa\sigma} &=& 
     4 (b_1+b_2-b_3-b_4-1) g^2 C a_\mu^a a^{\mu,a} \hat\sigma_0 
    + 4 (b_2 -b_3-b_4  ) g^2 C 
         d^{abh} a_\mu^a a^{\mu,b} \sigma^h \ .
   \ee
One can recombine the couplings so that
   \be
   {\cal L}_{vv\sigma} &=& 
    4 (\tilde b_1+\tilde b_2) C  v_\mu^a v^{\mu,a}\hat\sigma_0
    +  4 \tilde b_2 C d^{abh} v_\mu^a v^{\mu,b} \sigma^h \nn
   {\cal L}_{aa\sigma} &=& 
     4 (\tilde b_1+\tilde b_3-1) C a_\mu^a a^{\mu,a} \hat\sigma_0 
    + 4 \tilde b_3 C d^{abh} a_\mu^a a^{\mu,b} \sigma^h
   \ee
with the dimensionless couplings,
$\tilde b_1 \equiv b_1 g^2 $, $\tilde b_2 \equiv - (b_2+b_3+b_4) g^2 $ and
$\tilde b_3 \equiv  (b_2-b_3-b_4) g^2$.
The vector meson vertices $vvv$ and $vaa$ stem only from the
kinetic part of the vector meson Lagrangian
    \be
    {\cal L}_{vaa} &=& -4 g f^{abc}
    \left\{ v_\nu^a ( \partial^\mu a^{\nu,b}) a_\mu^c 
    + (\partial^\mu v_\nu^a ) a_\mu^b a^{\nu,c}
    + v_\mu^a A_\nu^b (\partial^\mu a^{\nu,c}) \right\}
    \nn
    {\cal L}_{vvv} &=& -4 g f^{abc} v_\nu^a ( \partial^\mu v^{\nu,b}) v_\mu^c .
    \ee
Here the three point vector vertex is fixed by the gauge coupling
$g$ and the structure constants.
The vector meson propagator is
    \be
    \Pi_v^{\mu\nu} &=& \frac{i}{q^2-m_v^2}
    \left( g^{\mu\nu}- \frac{q^\mu q^\nu}{m_v^2}\right) ,
    \ee
and similarly for the axial vector particles.

\subsection*{C-ii. Various Contributions}

The tree contribution to $\V\V\sigma_0$ is
    \be
    {\cal T}_{vv\sigma_0}^{ab} &=& 
     i\frac 83 (\tilde b_1+\tilde b_2)  \epsilon_1\cdot\epsilon_2
     \delta^{ab}\delta^{cd} \frac{C M_a}{E_a^2}
     \left(f_{v_c} f_{v_d} m_{v_c} m_{v_d}\right)
     \frac{i}{q_{v_c}^2-m_{v_c}^2}
     \frac{i}{s-m_{\sigma_0^2}} \langle 0 | \bar q q|\sigma\rangle
     \frac{i}{q_{v_c}^2-m_{v_c}^2}
    \ee
where $\langle 0 |\bar q q|\sigma\rangle =\lambda_0^2$. 
Since the photon is on mass-shell only the combination
$\epsilon_1\cdot\epsilon_2$ appear after contractions,
    \be
    \epsilon_1^\mu\Pi_v^{c,\mu\gamma}\Pi_v^{d,\gamma\nu}\epsilon_2^\nu
    &=& \epsilon_1\cdot\epsilon_2 
     \frac{i}{q_1^2-m_{v_c}^2} \frac{i}{q_2^2-m_{v_d}^2} \ .
    \ee
The tree contribution to $\V\V\sigma^h$ is
    \be
    {\cal T}_{vv\sigma^h}^{ab} &=& 
     -i4 \tilde b_2  \epsilon_1\cdot\epsilon_2
     d^{abh} d^{cdh} \frac{C M_b}{E_a E_b}
     \left(f_{v_c} f_{v_d} m_{v_c} m_{v_d}\right)
     \frac{i}{q_{v_c}^2-m_{v_c}^2}
     \frac{i}{s-m_{\sigma_0^2}} \langle 0 | \bar q \lambda^h q|\sigma^h\rangle
     \frac{i}{q_{v_c}^2-m_{v_c}^2}\ ,
    \ee
where $\langle 0 |\bar q \lambda^h q|\sigma^h \rangle =\lambda_h^2$.

The $\sigma_0$ contribution to $\V\V\j_A\j_A$ is
    \be
    {\cal T}_{vvaa}^{ab} &=& \delta^{cd}\delta^{ab}
    \epsilon_1\cdot\epsilon_2
    \frac{1}{E_a E_b} 
    \frac{i f_{v_c} m_{v_c}}{q_{v_c}^2-m_{v_c}^2}
    \frac{i f_{v_d} m_{v_d}}{q_{v_d}^2-m_{v_d}^2}
    \left( 16 (\tilde b_1+\tilde b_2) (\tilde b_1+\tilde b_3 -1) C^2\right)
    \frac{i}{s-m_{\sigma_0}^2}
    \nn &&
    \times (f_{a_a}m_{a_a}f_{a_b}m_{a_b})
    \times
    k_a^\alpha \Pi_a^{a,\alpha\gamma}\Pi_a^{a,\gamma\beta} k_b^\beta \ ,
    \ee
with
    \be
    k_1^\alpha \Pi_a^{a,\alpha\gamma}\Pi_a^{b,\gamma\beta} k_2^\beta
    &=&
    k_1\cdot k_2\left(1-\frac{k_1^2}{m_{a_a}^2}-\frac{k_2^2}{m_{a_b}^2}
       +\frac{k_1^2}{m_{a_a}^2}\frac{k_2^2}{m_{a_b}^2}\right)
     \frac{i}{k_1^2-m_{a_a}^2} \frac{i}{k_2^2-m_{a_b}^2} \ .
    \ee
The $\sigma_h$ contribution to $\V\V\j_A\j_A$ is
    \be
    {\cal T}_{vvaa}^{ab} &=& d^{cdh} d^{abh}
    \epsilon_1\cdot\epsilon_2
    \frac{1}{E_a E_b} 
    \frac{i f_{v_c} m_{v_c}}{q_{v_c}^2-m_{v_c}^2}
    \frac{i f_{v_d} m_{v_d}}{q_{v_d}^2-m_{v_d}^2}
    \left( 16 \tilde b_2 \tilde b_3 C^2\right)
    \frac{i}{s-m_{\sigma_h}^2}
    \nn &&
    \times (f_{a_a}m_{a_a}f_{a_b}m_{a_b})
    k_a^\alpha \Pi_a^{a,\alpha\gamma}\Pi_a^{b,\gamma\beta} k_b^\beta \ .
    \ee
In Fig.~\ref{fig:diag}-(b), the contribution from the intermediate
$\V$ vanishes because of the SU(3) structure constant $f^{abc}$.
The contribution from  Fig.~\ref{fig:diag}-(c) results in
    \be
    {\cal T}_{vvaa}^{ab} &=& i^2 f^{caf} f^{dbf} \frac{1}{E_a E_b}
      \left(\frac{i f_{v_c} m_{v_c}}{q_1^2-m_{v_c}^2}\right) 
      \left(\frac{i f_{v_d} m_{v_d}}{q_2^2-m_{v_d}^2}\right) 
      \left(\frac{i}{t-m_{a_f}^2}\right)
      \left(\frac{i f_{v_a} m_{a_a}}{k_1^2-m_{a_a}^2}\right) 
      \left(\frac{i f_{v_b} m_{a_b}}{k_2^2-m_{a_b}^2}\right) \nn
    && (-4 g)^2 \left[
      (\epsilon_1\cdot k_1) (\epsilon_2\cdot k_2)
         \left(1-\frac{(k_1-q_1)^2}{m_{a_f}^2}\right)
     \left( (k_1-q_1) (k_2-q_2) + k_1^2 +k_2^2
         -\frac{(k_1\cdot q_1)^2}{m_{a_a}^2}
         -\frac{(k_2\cdot q_2)^2}{m_{a_b}^2} \right) \right.  \nn 
    && \left.  +\left(\epsilon_1\cdot\epsilon_2
       +\frac{ (\epsilon_1\cdot k_1) (\epsilon_2\cdot k_2)}{m_{a_f}^2}\right)
      \left( k_1^2-\frac{(k_1\cdot q_1)^2}{m_{a_a}^2}\right)
      \left( k_2^2-\frac{(k_2\cdot q_2)^2}{m_{a_b}^2}\right) \right]\ .
    \ee
One can generally redefine the couplings,
     \be
     \tilde c_0 f_\pi       &\equiv& \frac 23 
                             (\tilde b_1+\tilde b_2) C \lambda_0^2 \nn
     \tilde c_{h=3,8} f_\pi &\equiv& - \tilde b_2 C 
                             \lambda_{h=3,8}^2 \nn
     \tilde g_1 f_\pi^2     &\equiv& (\tilde b_1 +\tilde b_2)
                               (\tilde b_1+\tilde b_3-1) C^2 \nn
     \tilde g_2 f_\pi^2     &\equiv& \tilde b_2 \tilde b_3 C^2 \nn
     \tilde g_3             &\equiv& (-g^2) 
     \ee
giving in total six independent parameters.

\subsection*{C-iii. Final Result}

$\sigma_0$ and $\sigma_h$ contributions to $\V\V\sigma$ :
     \be
     {\cal T}_{vv\sigma_0}^{ab} &=& 
       4 \tilde c_0 f_\pi  \epsilon_1\cdot\epsilon_2
      \delta^{ab}\delta^{cd} \frac{ M_a}{E_a^2}
      \frac{f_{v_c} f_{v_d}}{ m_{v_c} m_{v_d}}
      \frac{1}{s-m_{\sigma_0^2}} \nn
     {\cal T}_{vv\sigma^h}^{ab} &=& 
      4 \tilde c_h f_\pi \epsilon_1\cdot\epsilon_2
      d^{abh} d^{cdh} \frac{ M_b}{E_a E_b}
      \frac{f_{v_c} f_{v_d}}{m_{v_c} m_{v_d}}
      \frac{1}{s-m_{\sigma_0^2}} \ ,
     \ee
where the couplings $c_{0,8}$ will be fixed empirically.

$\sigma_0$ contribution to $\V\V\j_A\j_A$ :

    \be
    {\cal T}_{vvaa}^{ab} &=& i 16 \tilde g_1 f_\pi^2 \delta^{cd}\delta^{ab}
    \epsilon_1\cdot\epsilon_2 \frac{1}{E_a E_b} 
    \frac{f_{v_c} f_{v_d}}{m_{v_c} m_{v_d}} \frac{1}{s-m_{\sigma_0}^2}
    \nn && \times
    k_1\cdot k_2\left(1-\frac{m_a^2}{m_{a_a}^2}-\frac{m_b^2}{m_{a_b}^2}
       +\frac{m_a^2}{m_{a_a}^2}\frac{m_b^2}{m_{a_b}^2}\right)
      \frac{f_{a_a}m_{a_a}}{m_a^2-m_{a_a}^2} 
      \frac{f_{a_b}m_{a_b}}{m_b^2-m_{a_b}^2}\ .
    \ee

$\sigma_h$ contribution to $\V\V\j_A\j_A$ :

    \be
    {\cal T}_{vvaa}^{ab} &=& i 16\tilde  g_2 f_\pi^2 d^{cdh} d^{abh}
    \epsilon_1\cdot\epsilon_2
    \frac{1}{E_a E_b} 
    \frac{ f_{v_c} f_{v_d}}{m_{v_c} m_{v_d}}
    \frac{1}{s-m_{\sigma_h}^2}
    \nn &&
    \times
    k_1\cdot k_2\left(1-\frac{m_a^2}{m_{a_a}^2}-\frac{m_b^2}{m_{a_b}^2}
       +\frac{m_a^2}{m_{a_a}^2}\frac{m_b^2}{m_{a_b}^2}\right)
     \frac{f_{a_a}m_{a_a}}{m_a^2-m_{a_a}^2} 
     \frac{f_{a_b}m_{a_b}}{m_b^2-m_{a_b}^2}\ .
    \ee

Axial-vector contributions to $\V\V\j_A\j_A$ :

    \be
    {\cal T}_{vvaa}^{ab} &=& -i 16 g_3 f^{caf} f^{dbf}
      \frac{1}{E_a E_b}
      \left(\frac{f_{v_c} f_{v_d}}{m_{v_c} m_{v_d}}\right)
      \left(\frac{1}{t-m_{a_f}^2}\right)
      \left(\frac{ f_{v_a} m_{a_a}}{m_a^2-m_{a_a}^2}\right) 
      \left(\frac{ f_{v_b} m_{a_b}}{m_b^2-m_{a_b}^2}\right) \nn
    &&  \left[ (\epsilon_1\cdot k_1) (\epsilon_2\cdot k_2)
     \left(1-\frac{t}{m_{a_f}^2}\right) \left( -t + m_a^2 +m_b^2
     -\frac{(k_1\cdot q_1)^2}{m_{a_a}^2}
     -\frac{(k_2\cdot q_2)^2}{m_{a_b}^2} \right) \right.  \nn 
    && \left.
     +\left(\epsilon_1\cdot\epsilon_2
       +\frac{ (\epsilon_1\cdot k_1) (\epsilon_2\cdot k_2)}{m_{a_f}^2}\right)
      \left( m_a^2-\frac{(k_1\cdot q_1)^2}{m_{a_a}^2}\right)
      \left( m_b^2-\frac{(k_2\cdot q_2)^2}{m_{a_b}^2}\right)
     \right]\nn
    && + (t,a,k_1\leftrightarrow u,b,k_2).
    \ee
In the text, these couplings are redefined in terms of
dimensionless couplings.

\end{document}